\documentclass[11pt]{article}\textwidth 6.5in\textheight 9in
\usepackage{amssymb}\usepackage[colorlinks]{hyperref}\usepackage{color}
	\usepackage{stmaryrd}\usepackage{mathrsfs}
	\usepackage{graphicx}
	\usepackage{amsmath}
\usepackage{caption}

\begin{document}
\setlength{\captionmargin}{27pt}
\newcommand\hreff[1]{\href {http://#1} {\small http://#1}}
\newcommand\trm[1]{{\bf\em #1}} \newcommand\emm[1]{{\ensuremath{#1}}}
\newcommand\prf{\paragraph{Proof.}}\newcommand\qed{\hfill\emm\blacksquare}

\setcounter{tocdepth}{3} 

\newtheorem{thr}{Theorem} 
\newtheorem{lmm}{Lemma}
\newtheorem{cor}{Corollary}
\newtheorem{con}{Conjecture} 
\newtheorem{prp}{Proposition}

\newtheorem{blk}{Block}
\newtheorem{dff}{Definition}
\newtheorem{asm}{Assumption}
\newtheorem{rmk}{Remark}
\newtheorem{clm}{Claim}
\newtheorem{example}{Example}

\newcommand{\ab}{a\!b}
\newcommand{\yx}{y\!x}
\newcommand{\yux}{y\!\underline{x}}

\newcommand\floor[1]{{\lfloor#1\rfloor}}\newcommand\ceil[1]{{\lceil#1\rceil}}

\newcommand{\lea}{<^+}
\newcommand{\gea}{>^+}
\newcommand{\eqa}{=^+}

\newcommand{\lel}{<^{\log}}
\newcommand{\gel}{>^{\log}}
\newcommand{\eql}{=^{\log}}

\newcommand{\lem}{\stackrel{\ast}{<}}
\newcommand{\gem}{\stackrel{\ast}{>}}
\newcommand{\eqm}{\stackrel{\ast}{=}}

\newcommand\edf{{\,\stackrel{\mbox{\tiny def}}=\,}}
\newcommand\edl{{\,\stackrel{\mbox{\tiny def}}\leq\,}}
\newcommand\then{\Rightarrow}

\newcommand\km{{\mathbf {km}}}\renewcommand\t{{\mathbf {t}}}
\newcommand\KM{{\mathbf {KM}}}\newcommand\m{{\mathbf {m}}}
\newcommand\md{{\mathbf {m}_{\mathbf{d}}}}\newcommand\mT{{\mathbf {m}_{\mathbf{T}}}}
\newcommand\K{{\mathbf K}} \newcommand\I{{\mathbf I}}

\newcommand\II{\hat{\mathbf I}}
\newcommand\Kd{{\mathbf{Kd}}} \newcommand\KT{{\mathbf{KT}}} 
\renewcommand\d{{\mathbf d}} 
\newcommand\D{{\mathbf D}}

\newcommand\w{{\mathbf w}}
\newcommand\Ks{\mathbf{Ks}} \newcommand\q{{\mathbf q}}
\newcommand\E{{\mathbf E}} \newcommand\St{{\mathbf S}}
\newcommand\M{{\mathbf M}}\newcommand\Q{{\mathbf Q}}
\newcommand\ch{{\mathcal H}} \renewcommand\l{\tau}
\newcommand\tb{{\mathbf t}} \renewcommand\L{{\mathbf L}}
\newcommand\bb{{\mathbf {bb}}}\newcommand\Km{{\mathbf {Km}}}
\renewcommand\q{{\mathbf q}}\newcommand\J{{\mathbf J}}
\newcommand\z{\mathbf{z}}

\newcommand\B{\mathbf{bb}}\newcommand\f{\mathbf{f}}
\newcommand\hd{\mathbf{0'}} \newcommand\T{{\mathbf T}}
\newcommand\R{\mathbb{R}}\renewcommand\Q{\mathbb{Q}}
\newcommand\N{\mathbb{N}}\newcommand\BT{\{0,1\}}
\newcommand\FS{\BT^*}\newcommand\IS{\BT^\infty}
\newcommand\FIS{\BT^{*\infty}}\newcommand\C{\mathcal{L}}
\renewcommand\S{\mathcal{C}}\newcommand\ST{\mathcal{S}}
\newcommand\UM{\nu_0}\newcommand\EN{\mathcal{W}}

\newcommand{\supp}{\mathrm{Supp}}

\newcommand\lenum{\lbrack\!\lbrack}
\newcommand\renum{\rbrack\!\rbrack}

\renewcommand\qed{\hfill\emm\square}

\title{\vspace*{-3pc} Extending Chaitin's Incompleteness Theorem}

\author {Samuel Epstein\footnote{JP Theory Group. samepst@jptheorygroup.org}}

\maketitle
\begin{abstract}
Chaitin's incompleteness theorem states that sufficiently rich formal systems cannot prove lower bounds on Kolmogorov complexity. In this paper we extend this theorem by showing theories that prove the Kolmogorov complexity of a large (but finite) number of strings are inaccessible. This is done by first showing such theories have large information with the halting sequence. Then, by applying the independence postulate, such theories are shown to be inaccessible in the physical world.
\end{abstract}
\section{Introduction}

G\"{o}del's famous incompleteness theorem states that any theory  $\mathcal{F}$ that is consistent, recursively axiomatizable, and “sufficiently rich” (contains Robinson-arithmetic $\mathcal{Q}$, or $\mathcal{Q}$ can be interpreted in it) is incomplete, in that there exists true statements that cannot be proven in it.

It is well known that there is no recursive method to determine a non constant lower bound on Kolmogorov complexity, $\K$. Chaitin's incompleteness theorem proves there exist no logical means to prove lower bounds on $\K$. Let $\mathcal{F}$ be as above, and significantly strong to make assertions about the Kolmogorov complexity of strings. Furthermore, let $\mathcal{F}$ be sound. Then we get the celebrated theorem.\\

\noindent\textbf{Theorem. (Chaitin's Incompleteness Theorem)} \textit{For theory $\mathcal{F}$, there is a constant $c$ such that $\mathcal{F}$ does not prove $c<\K(x)$ for any $x$.}\\

The proof is straightforward. Assume otherwise. Take any $c$ and enumerate proofs of $\mathcal{F}$  until it proves the statement $c<\K(x)$ for some $x$. Then return $x$. This implies that $\K(x)<O(\log c)$, causing a contradiction for large enough $c$.

However this theorem doesn't prohibit the existence of formal systems that prove $c<\K(x)$ for a finite but very large number of strings. Or for our purposes, the above theorem doesn't prohibit theories which prove $\K(x)=c$ for a large (but finite) number of strings. Such theories are not to be expected to be accessible by logicians. In this paper, we prove such systems are exotic, and cannot exist in the physical world. To do so we use two steps. The first step proves the following theorem, which states $\K$ is uniformly uncomputable.\\

\noindent\textbf{Theorem.} \textit{A relation $X\subset\N\times \N$ of $2^n$ unique pairs $( b,\K(b))$ has $n\lel \I(X;\ch)$.}\\

The term $\ch$ is the halting sequence. The information term is $\I(x;\ch)=\K(x)-\K(x|\ch)$. The second part involves invoking the Indepedence Postulate (\textbf{IP}), introduced in \cite{Levin84,Levin13}. \textbf{IP}\;is an unprovable statement that physical sequences are independent from mathematical ones. Among other applications, \textbf{IP} can be interpreted as a finitary Church-Turing thesis. The statement is as follows.
\begin{quote}
\textbf{IP:}\textit{ Let $\alpha$ be a sequence defined with an $n$-bit mathematical statement (e.g., in Peano Arithmetic), and a sequence $\beta$ can be located in the physical world with a $k$-bit instruction set (e.g., ip-address). Then $\I(\alpha : \beta) < k+n+c$, for some small absolute constant $c$.}
\end{quote}
We rework \textbf{IP} so that $x=\alpha\in\FS$, $\beta$ is equal to the halting sequence $\ch$, and the information term is equal to $\I(x;\ch)=\K(x)-\K(x|\ch)$. Since $\ch$ can be described by an $O(1)$ bit mathematical sequence, we get
$$\I(x;\ch)\lea \mathbf{Address}(x).$$

Let $\mathcal{F}$ be a formal system defined in Chaitin's Incompleteness Theorem. Assume that $\mathcal{F}$ can be used to prove $\K(x_i)=c_i$ for $2^n$ unique strings $x_i$. Then by Theorem \ref{thr:main}, Lemma \ref{lmm:cons}, and \textbf{IP}, 
$$n\lel \I(\{(x_i,c_i)\};\ch)\lel \I(\mathcal{F};\ch)\lel \mathbf{Address}(\mathcal{F}).$$
Thus as the number strings with proved Kolmogorov complexities grows, the formal system $\mathcal{F}$  becomes exotic and by \textbf{IP}, inaccessible in the physical world. For related work, in \cite{Levin13}, it was shown that consistent completions of PA have infinite mutual information with $\ch$ and thus have infinite addresses. This paper extends this result by proving the existence of theories with finite mutual information with the halting sequence. Note that Theorem \ref{thr:main} can be generalized to binary relations that approximate Kolmogorov complexity.
\section{Conventions}
\label{sec:conv}
For positive real functions $f$, by ${\lea}f$, ${\gea}f$, ${\eqa}f$, and ${\lel} f$, ${\gel} f$, ${\sim} f$ we denote ${\leq}\,f{+}O(1)$, ${\geq}\,f{-}O(1)$, ${=}\,f{\pm}O(1)$ and ${\leq}\,f{+}O(\log(f{+}1))$, ${\geq}f\,{-}O(\log(f{+}1))$, ${=}\,f{\pm}O(\log(f{+}1))$. $\K(x|y)$ is the conditional prefix Kolmogorov complexity. The chain rule states $\K(x,y)\eqa \K(x) +\K(y|\K(x),x)$.  Let $[A]=1$ if the mathematical statement $A$ is true, otherwise $[A]=0$. Let $\K_t(x|y)=\inf\{\|p\|:U_y(p)=x \textrm{ in $t$ steps}\}$. The information the halting sequence $\ch$ has about $x$ is $\I(x;\ch|y)=\K(x|y)-\K(x|y,\ch)$. $\I(x;\ch)=\I(x;\ch|\emptyset)$. A probability measure is elementary if its support is finite and it has rational values. The deficiency of randomness of $x\in\FS$ with respect to elementary probability measure $Q$ is $\d(X|Q)=\ceil{-\log Q(X)-\K(x|\langle Q\rangle)}$. The stochasticity of $x$ is $\Ks(x)=\min_Q \K(Q)+3\log\max\{\d(X|Q),1\}$.
\begin{lmm}[\cite{Epstein21,Levin16}]
\label{lmm:ksh}
$\Ks(x)\lel\I(x;\ch)$.
\end{lmm}

\begin{lmm}[\cite{EpsteinDerandom22}]
\label{lmm:cons}For partial computable $f$, $\I(f(x):\ch)\lea \I(x;\ch)+\K(f)$.
\end{lmm}

\section{Results}

Let $\Omega = \sum\{2^{-\|p\|}:U(p)\textrm{ halts}\}$ be Chaitin's Omega, $\Omega_n\in\Q_{\geq 0}$ be be the rational formed from the first $n$ bits of $\Omega$, and $\Omega^t = \sum\{2^{-\|p\|}:U(p)\textrm{ halts in time $t$}\}$. For $n\in \N$, let $\bb(n) = \min \{ t : \Omega_n<\Omega^t\}$. $\bb^{-1}(m) = \arg\min_n \{\bb(n-1)<m\leq \bb(n)\}$. Let $\Omega[n]\in\FS$ be the first $n$ bits of $\Omega$.

\begin{lmm}
\label{lmm:rec}
For $n=\bb^{-1}(m)$, $\K(\Omega[n]|m,n)=O(1)$.
\end{lmm}
\begin{prf}
For a string $x$, let $BB(x) = \inf\{t:\Omega^t>0.x\}$. Enumerate strings of length $n$, starting with $0^n$, and return the first string $x$ such that $BB(x)\geq m$. This string $x$ is equal to $\Omega[n]$, otherwise let $y$ be the largest common prefix of $x$ and $\Omega[n]$. Thus $BB(y)=\bb(\|y\|)\geq BB(x)\geq m$, which means $\bb^{-1}(m)\leq \|y\|<n$, causing a contradiction.\qed
\end{prf}
$ $\newpage
\begin{thr}
\label{thr:main}
A relation $X\subset\N\times \N$ of $2^n$ unique pairs $( b,\K(b))$ has $n\lel \I(X;\ch)$.
\end{thr}
\begin{prf}
We relativize the universal Turing machine to $n$. 
Let $X=\{x_i,c_i\}_{i=1}^{2^n}$, and $T=\min\{t:\K_t(x_i)=c_i=\K(x_i),\textrm{ for }i=1,\dots,n\}$.  Let $N=\bb^{-1}(T)$ and $B=\bb(N)$. We relativize the universal Turing machine to $B$. Later on, we will make this relativization explicit. We also assume that $c_i>n$. If this is not the case, then one can construct $X'\subset X$ of size $2^{n-1}$ with $c_i>n-1$ and use $X'$ instead.

 Let $m(x) = 2^{-\K_B(x)}$. Let $Q$ be an elementary probability measure that realizes $\Ks(X)$ and $d=\max\{\d(X|Q),1\}$. Without loss of generality, the support of $Q$ is restricted to finite binary relations $B\subset \N \times \N$ of size $2^n$. Let $B_1=\bigcup\{y:(y,c)\in B\}$. Let $S=\bigcup\{B_1:B\in\mathrm{Support}(Q)\}$. We randomly select each string in $S$ to be in a set $R$ independently with probability $d2^{-n}$. Thus $\E[m(R)]\leq d2^{-n}$. For $B\in\mathrm{Support}(Q)$, 
\begin{align*}
&\E_R\E_{B\sim Q}[[R\cap B_1=\emptyset]]\\
=&\E_{B\sim Q}\Pr(R\cap B_1=\emptyset)\\
=& (1-d2^{-n})^{2^{n}} < e^{-d}.
\end{align*}
Thus there exists a set $R\subseteq S$ such that $\m(R)\leq 2{\cdot} 2^{-n}$ and $\E_{B\sim Q}[[R\cap B_1=\emptyset]] < 2e^{-d}$. Let $t(B)=.5[R\cap B_1=\emptyset]2^{d}$. $t$ is a $Q$-test, with $\E_{B\sim Q}[t(B)]\leq 1$. It must be that $t(X)\neq 0$, otherwise,
$$
1.44d-1 < \log t(X) \lea \d(X|Q)+\K(t|Q) \lea d+\K(d),
$$
which is a contradiction for large enough $d$, which one can assume without loss of generality. Thus $t(X)\neq 0$ and $R\cap X_1\neq\emptyset$. Furthermore, if $y\in R$, $\K(y)\lea -\log m(x)-n+\log d+\K(m,R)$. So for $x\in R\cap X_1$, making the relativization of $B$ explicit. 
\begin{align}
\nonumber
\K(x|B) & \lea -\log m(x)-n+\log d+\K(m,R|B)\\
\nonumber
\K(x)-\K(B)& \lea \K(x)-n+\log d+\K(S|B)\\
\nonumber
n &\lea \K(B)+\log d+\K(d,Q|B)\\
\nonumber
n &\lea \K(B)+\Ks(X|B)\\
\nonumber
n &\lel \K(B)+\I(X;\ch|B)\\
\label{eq:BH}
n &\lel \K(B)+\K(X|B)-\K(X|\ch)+O(\log N)
\end{align}
Equation \ref{eq:BH} is due to the fact that $B$ is computable from $\Omega[N]$, thus it is computable from $\ch$ and $N$.
So we have,
\begin{align}
\nonumber
&\K(X|B)+\K(B)\\
\nonumber
\lea& \K(X|B,\K(B))+\K(\K(B)|B)+\K(B)\\
\label{eq:chain}
\lea & \K(X,B)+\K(\K(B)|B)\\
\label{eq:NM}
\lea & \K(X,N,B)+O(\log N)\\
\label{eq:erM}
\lea & \K(X,N)+O(\log N).\\
\nonumber
\lea & \K(X)+O(\log N).\\
\label{eq1}
n \lel & \K(X)-\K(X|\ch)+O(\log N).
\end{align}
Equation \ref{eq:chain} is from the chain rule. Equation \ref{eq:NM} is from the fact that $M=\bb(N)$. Equation \ref{eq:erM} comes from $\K(T|X)=O(1)$ and Lemma \ref{lmm:rec}, which implies $\K(B|N,T) \lea \K(\Omega[N]|N,T) \lea  O(1)$. 

From $X$, one can compute $T$, where $\bb^{-1}(T)=N$. Therefore by Lemma \ref{lmm:rec}, $\K(\Omega[N]|X)\lea \K(N)$, so by Lemma \ref{lmm:cons}, 
\begin{align}
\label{eq2}
N&\lel \I(\Omega[N];\ch)\lel \I(X;\ch)+\K(N)\lel \I(X;\ch). 
\end{align}
The above equation used the common fact that the first $n$ bits of $\Omega$ had $n-O(\log n)$ bits of mutual information with $\ch$. 
So combining Equations \ref{eq1} and \ref{eq2}, we get
$$
n \lel\I(X;\ch).
$$
\qed
\end{prf}


\begin{thebibliography}{Eps22}

\bibitem[Eps21]{Epstein21}
Samuel Epstein.
\newblock All sampling methods produce outliers.
\newblock {\em IEEE Transactions on Information Theory}, 67(11):7568--7578,
  2021.

\bibitem[Eps22]{EpsteinDerandom22}
S.~Epstein.
\newblock {22 Examples of Solution Compression via Derandomization}.
\newblock {\em CoRR}, abs/2208.11562, 2022.

\bibitem[Lev84]{Levin84}
L.~A. Levin.
\newblock {Randomness conservation inequalities; information and independence
  in mathematical theories}.
\newblock {\em {Information and Control}}, 61(1):15--37, 1984.

\bibitem[Lev13]{Levin13}
L.~A. Levin.
\newblock Forbidden information.
\newblock {\em J. ACM}, 60(2), 2013.

\bibitem[Lev16]{Levin16}
L.~A. Levin.
\newblock Occam bound on lowest complexity of elements.
\newblock {\em Annals of Pure and Applied Logic}, 167(10):897--900, 2016.

\end{thebibliography}
\end{document}